\documentclass{ifacconf}

\usepackage{graphicx}      
\usepackage{natbib}        
\usepackage{units}
\usepackage{amsmath,amssymb,amsfonts}
\usepackage{tikz}
\usetikzlibrary{shapes, arrows.meta, positioning}
\usepackage{circuitikz} 
\usepackage{eurosym}
\usepackage{adjustbox}
\usepackage{multirow}
\usepackage{booktabs}
\usepackage{svg}
\usepackage{mathtools}

\newif\ifmargincomments 
\margincommentstrue

\ifmargincomments

\else

\fi

\begin{document}
\begin{frontmatter}

\title{Optimal Co-Design of a Hybrid Energy Storage System for Truck Charging\thanksref{footnoteinfo}} 

\thanks[footnoteinfo]{Research funded in part by the Ministry of Economic Affairs and Climate of The Netherlands}

\author[First]{Juan Pablo Bertucci} 
\author[First]{Sudarshan Raghuraman} 
\author[First]{Theo Hofman}
\author[First]{Mauro Salazar}

\address[First]{Eindhoven University of Technology, 
   Eindhoven, 5600 MB Netherlands (e-mail: j.p.bertucci@tue.nl).}

\begin{abstract}                
The major challenges to battery electric truck adoption are their high cost and grid congestion.
In this context, stationary energy storage systems can help mitigate both issues. Since their design and operation are strongly coupled, to make the best out of them, they should be jointly optimized.
This paper presents a co-design framework for hybrid energy storage systems where their technology and sizing are optimized jointly with their operational strategies.
Specifically, we consider a microgrid supporting truck chargers that consists of utility grid, solar panels, and energy storage systems including batteries, supercapacitors and flywheels.
We frame the co-design problem as a mixed-integer linear program that can be solved with global optimality guarantees.
We showcase our framework in a case-study of a distribution center in the Netherlands. Our results show that although the battery-only configuration is already competitive, adding supercapacitors or flywheel storage decrease total cost and increase energy sold back to the grid.
Overall, the fully hybrid solution (Battery+Supercapacitors+Flywheel) offers the best outcomes, achieving the lowest overall cost (1.96\% lower compared to battery-only) and reduced grid dependency, but at a higher (2.6\%) initial investment.
\end{abstract}


\end{frontmatter}

\section{Introduction}
The transportation sector is rapidly shifting towards vehicle electrification, which significantly reduces greenhouse gas emissions, lowers fuel costs, and improves air quality by replacing internal combustion engines with cleaner and more efficient power sources \citep{Etukudoh2024}.
Particularly significant are the emissions of heavy duty vehcles, that account for 28\% of all road transport emissions in the European Union, but are just 2\% of the vehicle park \citep{AddressingHeavydutyClimate2022}.
However, there are multiple barriers to the widespread adoption of Battery Electric Trucks (BETs): long charging times, high capital costs, and especially in places such as the Netherlands, grid-connection bottlenecks that limit the power needed for large-scale truck charging. 
Microgrids (MGs), equipped with Renewable Energy Sources (RES) and Energy Storage Systems (ESS), offer a promising avenue to address these challenges \citep{Chong2016}. 
MGs can operate either in conjunction with the main grid or completely islanded; they generate power locally (e.g., from solar or wind) and store energy in ESS units, thereby reducing reliance on the electrical grid.
For fast-charging stations, MGs help ensure a more stable supply and can contribute to overall grid stability through ancillary services such as voltage and frequency regulation \citep{Ding2015}.
Central to microgrid performance is the type of ESS employed. Battery storage systems (BSS) are the most common choice and are widely used for peak offsetting, with growing adoption in multiple countries \citep{Liu2020ESS}.
However, high-power-density options such as supercapacitors and flywheels offer quicker response times and can efficiently meet brief but intense power demands \citep{Chong2016}.
This could be especially useful for charging heavy-duty BET fleets, where individual charging sessions may require very high power. 
Given the increasing power ratings of fast chargers and the intermittent nature of renewable generation, a combination of energy and power dense storage solutions may prove advantageous for cost-effective and reliable microgrids.
\begin{figure}[!t]
    \centering
    \resizebox{\linewidth}{!}{%
        \input{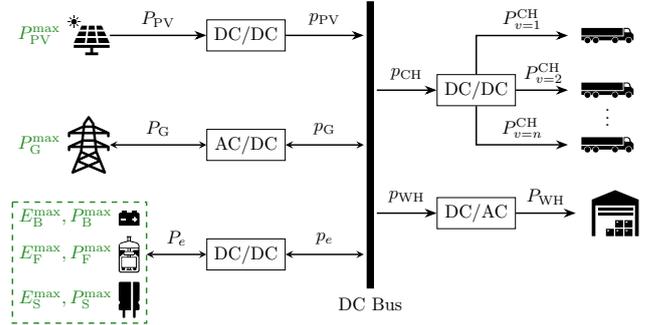}
    }
    \caption{Sketch of the microgrid and main problem variables. Design variables are indicated in green. Lowercase power variables are used for those directly connected to the DC bus.}
    \label{fig:1}
\end{figure}

\textit{Related Literature:}
Several works address the design and operation MGs. Most studies have focused on battery storage, examining how to size or schedule battery storage to provide load shifting and peak shaving \citep{Lin2003,Li2012}.
Research on supercapacitor storage has emphasized the high power density and rapid charge-discharge behavior \citep{Chong2016}, enabling MGs to handle fast, transient fluctuations in load or generation. Several studies have demonstrated that supercapacitor banks ranging from tens of kW to well over 1~MW can smooth short-term power spikes, improve stability, and costs \citep{Zhang2022,Song2020,Bian2019}. 
%
%
In parallel, flywheel-based energy storage systems offer advantages in frequency control and rapid power compensation \citep{FlywheelRef}. Flywheels can be sized from smaller units (100--250~kW) up to utility-scale devices approaching 1~MW or more \citep{Krein2020,Tudorache2019}. With round-trip efficiencies commonly around 85--95\%, flywheels excel in providing short bursts of power to maintain microgrid voltage and frequency within tight limits. Several projects document up to 20\% operational cost savings by offsetting high demand charges and reducing battery cycling while also offering high cycle life and minimal maintenance~\citep{Ortega2018}.
%
%
From a methodological standpoint, mathematical optimization is well established tool to tackle MG design problems. Mixed-Integer Nonlinear Programming (MINLP) was used in \citep{Ding2015} to optimize fast-charging stations integrated with ESS, although no renewable generation was included. A related study \citep{Liu2020ESS} employed Mixed Integer Liner Programming (MILP) to size PV and battery systems for EV charging but considered normally distributed arrivals of standard vehicles. 
\\
%
%
\textit{Statement of Contributions:}
Although many studies explore MG design with mathematical optimization, none to date perform co-design (joint optimization of plant and control decisions) of a MG with HSS consisting of batteries, supercapacitors, and flywheels in a single framework to serve large-scale BET charging at warehouses for realistic schedules. This study addresses this research gap.
\\
\textit{Organization:} 
The rest of the paper is structured as follows: Section~\ref{sec:method} presents the problem formulation and optimization model. Section~\ref{sec:results} discusses the results for our case study, and Section~\ref{sec:conclusion} concludes the paper.
\section{Methodology}\label{sec:method}
This section describes the co-design methodology. We outline the data inputs, the main equations that build the optimization problem, and obtention of a representative dataset for the design.
\subsection{Definitions and Data Inputs}
Our MG is analyzed over a simulation period \( T \) in days, which we discretize into intervals of size \( \tau \) minutes, resulting in a total of \( K = \frac{T}{\tau} \) time steps. Thus, our optimization will be performed over a discrete time horizon $k \in \mathcal{K} = \{0,1,2,\dots,K\}$ .
The MG incorporates multiple energy sources, where the set of energy sources is given by $s \in \mathcal{S} = \{\text{PV}, \text{G}\}$ where $\mathcal{S}$ represents the available energy sources, consisting of photovoltaic ($\mathrm{PV}$) generation and the utility grid ($\mathrm{G}$).
Each energy source $s$ generates power at time step $k$, denoted as $P_{s,k}$ which is the power supplied by source $s$ at time $k$. The efficiency factor of each energy source is denoted by $\eta_{\mathrm{s}}$ which accounts for conversion losses associated with each source. 
The MG includes multiple ESSs, represented by the set $e \in \mathcal{E} = \{\text{B}, \text{S}, \text{F}\}$ where $\mathcal{E}$ consists of Battery (B), Supercapacitors (S), and Flywheels (F) ESSs respectively.
The stored energy in ESS $e$ at time $k$ is given by $E_{e,k}$ while the power charged or discharged from the ESS is represented as $P_{e,k}$, and the charging and discharging efficiencies of each storage system are denoted by $\eta_{e,\mathrm{c}}$, $\eta_{e,\mathrm{d}}$ respectively.
The set of energy demand categories $\mathcal{D}$ is defined as $d \in \mathcal{D} = \{\text{CH}, \text{WH}\}$ which includes EV charging loads and power consumption from the Warehouse (WH) respectively.
The power demand from load category $d$ at time step $k$ is given by $P_{d,k}$, and the efficiency of power conversion for each demand category is denoted by $\eta_{\mathrm{d}}$.
The costs for deploying an energy storage system \(e\) can be determined from either a power-dependent or an energy-dependent component, where $C_{e}^{\mathrm{P}} (\text{\euro/kW})$ and $C_{e}^{\mathrm{E}} (\text{\euro/kWh})$ are the per-unit power capacity and per-unit energy cost, respectively.
Similarly, for the energy sources \(s\), the cost per unit \(\mathrm{kW}\) of installed PV is given by \(C_{\mathrm{PV}} \,(\text{\euro/kW})\). The cost of electricity purchased from the grid at time step \(k\) is represented by \(C_{\mathrm{G},k}\,(\text{\euro/kWh})\), which varies according to the market. Additionally, the MG incurs a fixed monthly grid connection cost, denoted as \(C^{\mathrm{conn}}_{\mathrm{G}}\) \((\text{\euro/kW})\), which depends on the contracted grid power capacity.
%
%
%
\subsection{Decision Variables}
The power flow within the MG is controlled through a set of decision variables that govern energy generation, storage, and distribution. 
The notation follows the one given in Fig.~\ref{fig:1}.
The power injected into the system from an energy source $s$ at time step $k$ is denoted by uppercase $P_{s,k}$ while the power supplied from an ESS $e$ at time step $k$ is given by $P_{e,k}$. 
In the case of grid power,  $P_{\mathrm{G},k}^{\mathrm{-}}$ represents power fed into the grid, while $P_{\mathrm{G},k}^{\mathrm{+}}$ represents power drawn from the grid.
Similarly, the power delivered to the demand category $d$ at time step $k$ is represented as $P_{d,k}$. 
The charge and discharge states of an ESS $e$ are repsented by $P_{e,k}^{\mathrm{+}}$ (power delivered during discharge) and $P_{e,k}^{\mathrm{-}}$(power drawn during charging). 
The installed/contracted capacity of an energy source $s$ is represented by $P_s^{\mathrm{\max}}$, while the maximum power capacity of an ESS $e$ is given by $P_e^{\mathrm{\max}}$. 
The total energy storage capacity of ESS $e$ is denoted as $E_e^{\mathrm{\max}}$.
Lowercase $p$ represents power adjusted for efficiency, accounting for conversion losses - for example $p_{s,k}$ and $p_{e,k} $.
\subsection{Constraints}
The optimization problem is subject to a set of constraints that govern the operation of the MG components, ensuring energy balance, operational limits, and proper switching behavior:
\subsubsection{Energy Sources}
The PV power reaching the DC bus $p_{\mathrm{PV},k}$ is given by:
\begin{equation}
    \label{eq:1}
    p_{\mathrm{PV},k} = \eta_{\mathrm{PV}} \, P_{\mathrm{PV},k},
\end{equation}
whereas the grid power flow can be bidirectional, which means power can be both imported or exported. 
The net grid power is simply the difference between the power drawn from the grid and power fed into it:
\begin{equation}
    p_{{\mathrm{G}},k} =  p_{{\mathrm{G}},k}^{\mathrm{+}} - p_{{\mathrm{G}},k}^{\mathrm{-}}. \label{eq:pm}
\end{equation}
The grid power exported ($p_{{\mathrm{G}},k}^{\mathrm{-}}$) and imported ($p_{{\mathrm{G}},k}^{\mathrm{+}} $) to and from the DC bus are defined as: 
%
\begin{equation}
  p_{\mathrm{G},k}^{-}
  = \frac{P_{\mathrm{G},k}^{-}}{\eta^{\mathrm{d}}_{\mathrm{G}}},
  \quad
  p_{\mathrm{G},k}^{+}
  = \eta^{\mathrm{c}}_{\mathrm{G}}\,P_{\mathrm{G},k}^{+}.
\end{equation}
Like grid power, ESS power can be positive when discharging and negative when charging.  
The net ESS power is difference between the power discharged and power charged. 
\begin{equation}
    p_{e,k} =  p_{e,k}^{\mathrm{+}} - p_{e,k}^{\mathrm{-}}
\end{equation}
The ESS charge power ($p_{e,k}^{\mathrm{-}}$) and discharge power ($p_{e,k}^{\mathrm{+}}$) flow to and from the DC bus, respectively. 
\begin{equation}
    p_{e,k}^{\mathrm{-}} = \frac{P_{e,k}^{\mathrm{-}}}{\eta^{\mathrm{c}}_{e}},\quad
    p_{e,k}^{\mathrm{+}} = \eta^{\mathrm{d}}_{e} P_{e,k}^{\mathrm{+}}.
\end{equation}

%
%
%
The total power supplied to the DC bus at each time step \(k\) must be equal to the total demand at the DC:
\begin{equation}
    \Bigl(\sum_{e \in \mathcal{E}} p_{e,k} \;+\; p_{\mathrm{PV},k} \;+\; p_{\mathrm{G},k}\Bigr)\
    \;=\; \sum_{d \in \mathcal{D}} p_{d,k}.
\end{equation}
The installed capacities of the sources and storage devices are constrained by maximum capacities due to space and technical constraints $P_{s}^{\mathrm{M}}$,$P_{e}^{\mathrm{M}}$ and $E_{e}^{\mathrm{M}}$ through:
\begin{equation}
    0 \;\leq\; P_{s}^{\mathrm{\max}} \;\leq\; P_{s}^{\mathrm{M}}, \quad \forall s\in\mathcal{S}
\end{equation}
\begin{equation}
    0 \;\leq\; P_{e}^{\mathrm{\max}} \;\leq\; P_{e}^{\mathrm{M}}, \quad \forall e \in \mathcal{E}
\end{equation}
\begin{equation}
    0 \;\leq\; E_{e}^{\mathrm{\max}} \;\leq\;E_{e}^{\mathrm{M}}, \quad \forall e \in \mathcal{E}. 
\end{equation}
Subsequently, the operational limits of power and energy values are bounded by:
\begin{equation}
    -\,P_{\mathrm{G}}^{\mathrm{\max}} \;\leq\; p_{\mathrm{G},k} \;\leq\; P_{\mathrm{G}}^{\mathrm{\max}}, \quad \forall s \in \mathcal{S},\quad \forall k \in \mathcal{K},
\end{equation}
\begin{equation}
    0 \;\leq\; P_{k}^{\mathrm{PV}} \;\leq\; p_{\mathrm{PV}}^{\mathrm{\max}},\quad \forall s \in \mathcal{S}, \quad \forall k \in \mathcal{K},
\end{equation}
\begin{equation}
    -\,P_{e}^{\mathrm{\max}} \;\leq\; p_{e,k} \;\leq\; P_{e}^{\mathrm{\max}}, \quad \forall e \in \mathcal{E}, \quad \forall k \in \mathcal{K},
\end{equation}
\begin{equation}
    0 \;\leq\; E_{e,k} \;\leq\; E_{e}^{\mathrm{\max}}, \quad \forall e \in \mathcal{E}, \quad \forall k \in \mathcal{K}.
\end{equation}
\subsubsection{Energy Storage System Dynamics}
The evolution of the state of energy for each ESS $e \in \mathcal{E}$ is governed by:
\begin{equation}
    E_{e,k+1} = E_{e,k} + B_e \cdot p_{e,k}^{\mathrm{+}} + C_e p_{e,k}^{\mathrm{-}} \quad \forall e \in \mathcal{E}, \quad \forall k \in \mathcal{K}.
\end{equation}
Here,  $B_e = -\frac{\tau}{\eta_{e}^{\mathrm{d}}}$ represents the contribution of power input/output to the SoE based on the discharge efficiency $\eta_{e}^{\mathrm{d}}$, while $C_e = \tau  \cdot \eta_{e}^{\mathrm{c}}$ accounts for charging efficiency $\eta_{e}^{\mathrm{c}}$. 

%
%
%
A limit is set on the depth of discharge for each energy storage device through: 
\begin{equation}
    E_{e,k} \;\geq\; E^\mathrm{\min}_{e}, \quad e \in \mathcal{E}.
\end{equation}
where $E^\mathrm{\min}_{e} = DoD \cdot E_{e}^{\mathrm{\max}}$ 
Additionally, a periodicity constraint for the storage is enforced as:
\begin{equation}
    E_{e,K} \;\geq\; E_{e,1}, \quad e \in \mathcal{E}.
\end{equation}

We define the C-rate $R_{e,k}$ of storage \( e \) over interval \( k \) as: 
\begin{equation}
    R_{e,k} = \frac{\Bigl| E_{e,k+1} - E_{e,k} \Bigr|}{E^{\mathrm{\max}}_{e}}, \quad k \in \mathcal{K},\quad \forall\, e \in \mathcal{E}.
    \label{eq:crate_def}
\end{equation}
We limit the $R$ of each storage according to their technical capabilities and what is allowed by degradation considerations. 
\begin{equation}
    R_{e,k} \leq R^\mathrm{M}_e, \quad \forall\, e \in \mathcal{E},\; k \in \mathcal{K}.
    \label{eq:crate_limit}
\end{equation}
To keep the formulation linear, we define an auxiliary variable
\begin{equation}
    q_{e,k} = E_{e}^\mathrm{\max}\cdot R_{e,k},
    \label{eq:q_def}
\end{equation}
so that by eq.~\eqref{eq:crate_def}, we have
\begin{equation}
    q_{e,k} = \Bigl| E_{e,k+1} - E_{e,k} \Bigr|.
    \label{eq:consistency}
\end{equation}
And the bilinear term in eq.(\ref{eq:q_def}) is linearized by the following constraints on $q_{e,k}$,
\begin{align}
    q_{e,k} &\geq 0, \label{eq:mcc1}\\
    q_{e,k} &\geq E^\mathrm{M}_e\, R_{e,k} + E^{\mathrm{max}}_{e}\, R^\mathrm{M}_e - E^\mathrm{M}_e\, R^\mathrm{M}_e, \label{eq:mcc2}\\
    q_{e,k} &\leq E^\mathrm{M}_e\, R_{e,k}, \label{eq:mcc3}\\
    q_{e,k} &\leq E^\mathrm{\max}_{e}\, R^\mathrm{M}_e. \label{eq:mcc4}
\end{align}

\subsection{Objective Function}
The objective of the co-design algorithm is to minimize the total cost of ownership (\(J^\mathrm{tot}\)), which is composed of operational (\(J^\mathrm{op}\)) and capital (\(J^\mathrm{cap}\)) costs.

\noindent
\textit{Operational Cost.} The operational expenditure \(J^\mathrm{op}\) is the agreggation of yearly operational costs \(J^\mathrm{op}_{y}\), which include energy costs from the grid, and maintenance of both storage and sources. 
For a year $y$:
\begin{equation}
    J^\mathrm{op}_y
    \;=\; \sum_{e}J^\mathrm{op}_{e} \;+\; \sum_{s}J^\mathrm{op}_{s}.
\end{equation}
For the energy storage sources $e$, we have fixed $C^{\mathrm{P,OM}}_{e}$ and variable $C^{\mathrm{E,OM}}_{e}$ Operation and Mantainance (OM) costs. 
Fixed costs are considered proportional to the maximum power $P^{\mathrm{\max}}_{e}$ of each power source, and variable costs proportional to the total energy throughput of each storage $Q_e$, which we define:
\begin{equation}
    Q_e
    \;=\; \sum_{k=1}^{k^\mathrm{sim}}q_{e,k}.
\end{equation}
So the operational expenditures (OpEx) can be written as:
\begin{equation}
    J^\mathrm{op}_{e} = C^{\mathrm{P,OM}}_{e} \cdot P^{\mathrm{\max}}_{e} + C^{\mathrm{E,OM}}_{e} \cdot Q_e
\end{equation}
Similarly, for the power sources $s$, the OpEx is given by adding the net cost of the energy purchased and sold from the power grid during the period of analysis, the connection costs, and OM costs for PV:
\begin{equation}
    J^\mathrm{op}_{s} 
    \;=\; \tau \cdot \sum_{k} c_{\mathrm{G},k}\cdot P_{\mathrm{G},k} 
      +J^{\mathrm{connec}}_{\mathrm{G}} +
     C^{\mathrm{OM}}_\mathrm{PV}\cdot P^{\mathrm{\max}}_{\mathrm{PV}}
\end{equation}
where the cost of energy from the grid can be positive or negative depending on the sign, and is pondered by the $f_{\mathrm{sell}}$ factor:
\begin{equation}
    c_{\mathrm{G},k} = -f_{\mathrm{sell}}\cdot C_{\mathrm{G},k} \cdot P_{\mathrm{G},k}^{\mathrm{-}} + C_{\mathrm{G},k} \cdot P_{\mathrm{G},k}^{\mathrm{+}}. 
\end{equation}
The total grid connection cost ($J^{\mathrm{connec}}_{\mathrm{G}}$) is given by :
\begin{equation}
    J^{\mathrm{connec}}_{\mathrm{G}}=  C^{\mathrm{conn}}_{\mathrm{G}} + C^{\mathrm{tran}}_{\mathrm{G}} + C^{\mathrm{var}}_{\mathrm{G}} \cdot P_{\mathrm{G}}^{\mathrm{\max}} + C^{\mathrm{peak}}_{\mathrm{G}} \cdot P_{\mathrm{G}}^{\mathrm{peak}}
\end{equation}
where $^{\mathrm{conn}}_{\mathrm{G}}$ is Fixed grid connection Cost, $C^{\mathrm{tran}}_{\mathrm{G}}$ is fixed transmission cost, $C^{\mathrm{var}}_{\mathrm{G}}$ is the variable grid connection cost based on the $P_{\mathrm{G}}^{\mathrm{\max}}$ and $C^{\mathrm{peak}}_{\mathrm{G}}$ is the cost associated with delivering peak power $P_{\mathrm{G}}^{\mathrm{peak}}$. 
We use the net present value formula to account for the discounted value of future revenues and costs, using a disount rate $r$:
\begin{equation}
    J^\mathrm{op}
    \;=\; \sum_{y=1}^{Y} \frac{1}{(1+r)^y} \cdot J^\mathrm{op}_{y}.
\end{equation}
\noindent
\textit{Capital Costs}. The total capital cost consists of the investments in the energy storage system, PV generation, and grid connection, which can be written as:
\begin{equation}
J^{\mathrm{cap}} 
~=~ J^{\mathrm{cap}}_{e} + J^{\mathrm{cap}}_{s},
\end{equation}
where
\begin{equation}
    J^{\mathrm{cap}}_{e} 
    \;=\; \max
        \Bigl(E_e^{\mathrm{\max}}\,C_e^{\mathrm{E}} \;,\; P_e^{\mathrm{\max}}\,C_e^{\mathrm{P}} \Bigr), \forall e \in \mathcal{E}
\end{equation}
\begin{equation}
    J^{\mathrm{cap}}_{s} 
    \;=\; P_{\mathrm{PV}}^{\mathrm{\max}} \, C_{\mathrm{PV}}.
\end{equation}
Additionally, we consider the resale value of the equipment at the end of life, proportional to the relative number of cycles remaining:
\begin{equation}
f_{\mathrm{re}} \cdot \frac{N^{\mathrm{rem}}_{\mathrm{c}}} {N_{\mathrm{c}}} \cdot E^{\mathrm{\max}}_e ,
\end{equation}
where $f_{\mathrm{re}}$ is the resale factor for the equipment. Considering we can re write the term $N^{\mathrm{rem}}_{\mathrm{c}}\cdot E^{\mathrm{max}}_e $ as:
$N^{\mathrm{rem}}_{\mathrm{c}} \cdot E^{\mathrm{\max}}_e = N_{\mathrm{c}} \cdot E^{\mathrm{\max}}_e - Q_e$
this, pondered by the discount factor at year $Y$, we have our final resale value:
\begin{equation}
    \label{eq:prelast}
    J^\mathrm{res}
    \;=\; \frac{1}{(1+r)^Y} \cdot f_{\mathrm{re}} \cdot \Bigl( \frac{  N_{\mathrm{c}} \cdot E^{\mathrm{\max}}_e - Q_e} {N_{\mathrm{c}}}  + J^{\mathrm{cap}}_{s} \Bigr)
\end{equation}
Accordingly, the total cost of ownership is reformulated as
\begin{equation}
    \label{eq:last}
    J = J^\mathrm{cap} + J^\mathrm{op} - J^\mathrm{res},
\end{equation}
where $J^\mathrm{cap}$ are the capital costs, $J_{w}^\mathrm{op}$ are the operational costs computed over the representative day $w$, and $J^\mathrm{res}$ are the final resale values of the amortized equipment.
%
\subsection{Optimal Co-design Problem}
We can thus write the minimization of the MG's cost as:
\noindent
\textbf{Problem 1} (Joint MG Design and Control)\\
\textit{Given a set of power demands \(P_d\), energy prices \(C_\mathrm{G}\), and equipment costs \(C_s\), the optimal sizing of storage \(E^{\mathrm{max}}_s\), contracted grid capacity \(P^{\mathrm{max}}_s\) and power dispatch \(P_{e,k},P_{s,k}\) is found by solving:}
\begin{equation*}
\begin{aligned}
\label{eq:final-prob}
\min_{P_{e,k},P_{s,k},E^{\mathrm{\max}}_s} \quad & J\bigl(P_{e,k},P_{s,k},E^{\mathrm{\max}}_s\bigr) \\
\textrm{s.t.} \quad 
& \eqref{eq:1}-~\eqref{eq:last}\\
& P_{e,k},P_{s,k},E^{\mathrm{\max}}_s \in \mathbb{R}
\end{aligned}
\end{equation*}
Problem~1 is a mixed-integer linear program that can be solved with global optimality guarantees. Considering the objective of minimizing cost, the convex relaxation~\eqref{eq:pm} is lossless (i.e., complementarity between positive and negative power components) for any $f_{\mathrm{sell}}\in(0,1]$.

\subsection{Synthetic Scenario Generation}
Solving the full-scale MG design and control problem over a 20-year horizon at a high temporal resolution is computationally intractable.
In practice, the Linear Programming is solved over a a few representative days at a high resolution and the results are extrapolated to capture the long-term performance.

To this end, we generate a set of representative operational days $T_\mathrm{syn}$ that capture the average behaviour and variability in solar irradiance, power demand, and energy prices.
Let $\mathbf{x}_i \in \mathbb{R}^\ell$, for $i=1,\ldots,N_\mathrm{hist}$, denote the vector of $\ell$ statistical moments of $P_{\mathrm{PV}}$, $P_{\mathrm{d}}$, and $C_{\mathrm{G}}$; for each day $i$ in a historical dataset.
We partition these days into $W$ clusters by solving the k-means problem:
\begin{equation}
\min_{\{z_i\}, \{c_w\}} \quad \sum_{i=1}^{N_\mathrm{hist}} \left\| \mathbf{x}_i - c_{z_i} \right\|^2, \quad z_i \in \{1,2,\dots,W\},
\label{eq:kmeans}
\end{equation}
where $c_w \in \mathbb{R}^\ell$ is the centroid of cluster $w$.
For each cluster $w$, we define the representative day as the day closest to the centroid:
\begin{equation}
\tilde{i}_w = \arg\min_{i\,:\,z_i = w} \left\| \mathbf{x}_i - c_w \right\|.
\label{eq:rep_day}
\end{equation}
The probability associated with the representative day for cluster $w$ is given by
\begin{equation}
\pi_w = \frac{\bigl|\{i\,:\,z_i = w\}\bigr|}{N_\mathrm{hist}}.
\label{eq:prob_day}
\end{equation}
With this, the number of days corresponding to each cluster is given by $T_{\mathrm{syn}}\cdot\pi_w$. 
To ensure that challenging operating conditions are not overlooked all clusters are represented in at least one day of data.
The sequence of the representative days is then generated by sampling the Markov transition matrix $T\in\mathbb{R}^{|W| \times |W|}$ from the historical sequence of cluster labels. 
This is obtained by counting through a frequentist approach: given the number of transitions between clusters
\[
n_{ij} = \#\{\,i: z_i=i \mbox{ and } z_{i+1}=j\,\},
\]
then we can obtain the transition matrix $A_{ij}$ by normalizing the counts
\begin{equation}
A_{ij} = \frac{n_{ij}}{\sum_{j'=1}^k n_{ij'}}, \quad i,j=1,\dots,k.
\label{eq:transition}
\end{equation}
With this, we obtain a synthetic sequence of $T_{\mathrm{syn}}$ days in a sequence given by the Markov chain defined by $A_{ij}$.
%


%


\section{Results}\label{sec:results}
In this section we describe the input parameters and experiments conducted to test the co-design algorithm. 
Table~\ref{tab:t1} summarizes the main parameter values used.
Historical values for solar power generation $P_\mathrm{PV}$ are obtained from \citep{PVGIS} and callibrated with local power generation data from project partners.
Prices for energy are obtained from \citep{ENTSOE_Transparency}.
The charging loads $P_{\mathrm{d}}$ are obtained from the results of \citep{JP} and combined with real distribution center data from project partners.
The total number of days for the historical data analysis $N_{hist}$ spans from 1/1/2021 to 31/12/2023.
\begin{table}[tbp]
  \centering
  \caption{Model parameters used for the case study.}
  \label{tab:t1}
  \renewcommand{\arraystretch}{1.2} 
  \begin{adjustbox}{max width=\linewidth}
    \begin{tabular}{lll|lll}
      \toprule
      \textbf{Var.} & \textbf{Value} & \textbf{Unit} & \textbf{Var.} & \textbf{Value} & \textbf{Unit} \\
      \midrule
      $\tau$ & 60   & min   & $Y$ & 20   & years \\
      $r$    & 0.04  & -     & $T$ & 30   & days \\
      \midrule
      $\eta_\mathrm{G}^\mathrm{c}$ & 0.95  & -    & $\eta_\mathrm{G}^\mathrm{d}$ & 0.95 & - \\
      $\eta_\mathrm{PV}$  & 0.9  & -    & $\eta_e^\mathrm{c}$  & [0.83, 0.95, 0.85] & - \\
      $\eta_e^\mathrm{d}$ & [0.88, 0.97, 0.93] & -  & $N_{\mathrm{c}}$            & [5,500,100]   & $1000 \cdot$ cycles  \\
      \midrule
      $C^\mathrm{P}_e$              & [1590,350,300] & k\euro/MW             & $C^\mathrm{P}_e$                   & 0.001          & - \\
      $C_\mathrm{PV}^{\mathrm{P}}$   & 300            & k\euro/MW      & $C_\mathrm{PV}^{\mathrm{OM}}$        & 15             & k\euro/MW \\
      $C^\mathrm{op}$              & 15             & k\euro/kWh     & $C^{\mathrm{E,OM}}_{e}$             & [3,5,10]       & k\euro/MWh \\
      $C^{\mathrm{P,OM}}_{e}$       & [30,6,20]      & k\euro/kW      & $C_{e}^{\mathrm{E}}$               & [900,1150,3000] & k\euro/MWh \\
      $P_{e}^{\mathrm{M}}$          & [10,30,20]     & k\euro         & $C_\mathrm{G}^\mathrm{peak}$       & 9.03           & k\euro/MW \\
      $f_\mathrm{PV}$             & 0.75           &  -      & $f_\mathrm{e}$       & 0.85           & -\\
      \midrule
      $P_{s}^{\mathrm{M}}$          & [2.8,5]        & MW             & $R^{\mathrm{M}}_{e}$               &  [3,100,10]   &  \\
      \midrule
      $E_{e}^{\mathrm{M}}$          & [5,0.2,0.5]    & MWh            & $E^{\mathrm{min}}_e$                & [0.15,0,0]    & \% \\
      \bottomrule
    \end{tabular}
  \end{adjustbox}
  \raggedright
  Notes: Values adapted from \cite{minear2018,mongird2019,kebede2022,IRENA2024}. $\mathcal{S} = \{\mathrm{PV},\mathrm{G}\}$, $\mathcal{E} = \{\mathrm{B},\mathrm{S},\mathrm{F}\}$
\end{table}
We used Yalmip (\cite{Lofberg2004}) to formulate Problem 1 and employed Gurobi (\cite{gurobi}) as the solver.
Solving the problem to a gap of less than 1\% took on average 8 minutes to parse and 2 minutes to solve on an Intel Core i7-12700H, 2.30 GHz processor with 16 GB RAM.
\subsubsection{Optimization Period Synthesis}
\begin{figure}[!h]
  \centering
  \resizebox{\linewidth}{!}{%
      \includegraphics{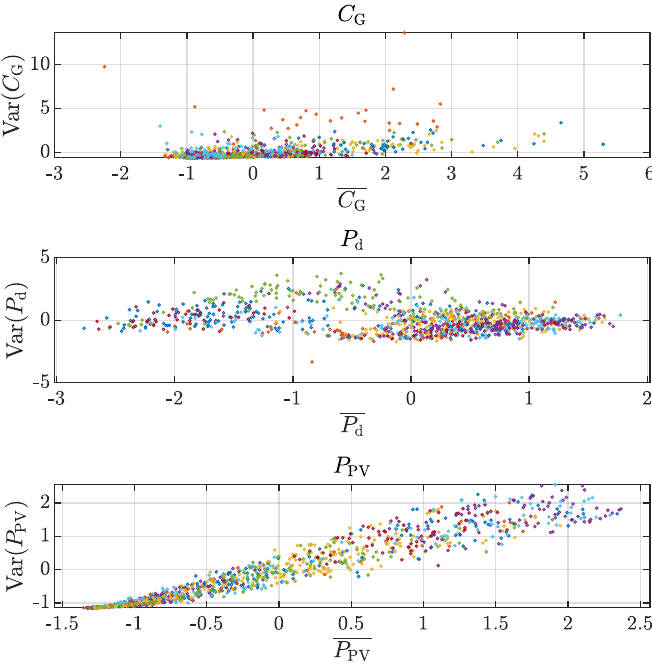}
  }
  \caption{Obtained cluster groups for grid prices $C_\mathrm{G}$, power demand $P_{\mathrm{d}}$ and solar power production $P_{\mathrm{PV}}$,with each colored dot representing a different cluster}
  \label{fig:clusters}
 
\end{figure}

The K-means clustering results for the time period are shown in Fig.\ref{fig:clusters}. 
20 clusters were used ($W=20$), from which a full month was created for $T_{\mathrm{syn}}$.
\subsubsection{Design Results}
\begin{figure}[!t]
  \centering
  \resizebox{\linewidth}{!}{%
  \def\svgwidth{0.8\textwidth}
      \includegraphics{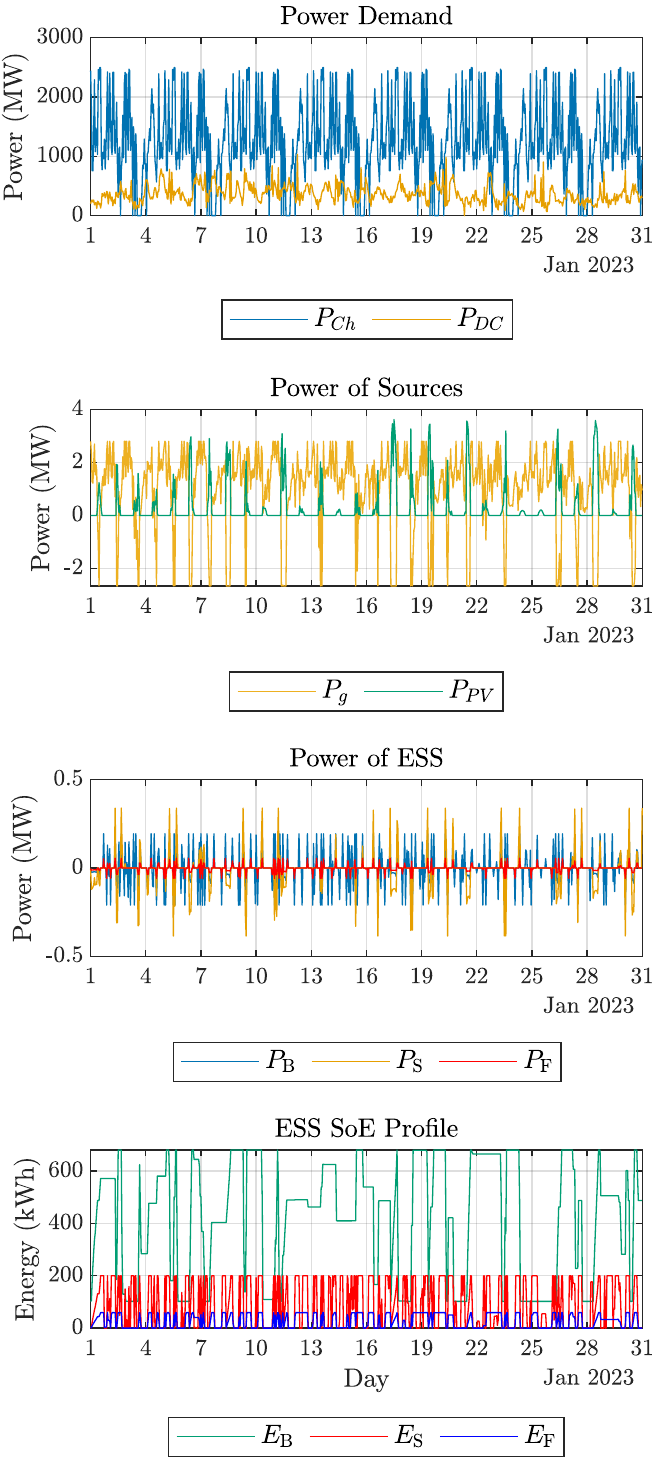}
  }
  \caption{Power Demands $P_{\mathrm{d}}$, Sources $P_{s}$, Storage $P_{e}$, and Energy $E_{e}$ for Exp.\ 4.}
  \label{fig:powers}
\end{figure}

Fig.~\ref{fig:powers} presents the operation curves for the final experiment (Experiment~4) with all ESS units. 
The charging demands $P_{\mathrm{CH}}$ are significantly higher than exsting warehouse consumption $P_{\mathrm{WH}}$
We observe the complementarity between solar $P_{\mathrm{PV}}$ and grid power $P_{\mathrm{G}}$ .
The battery operates on a longer time scale, handling bulk energy shifting, often during high prices or low solar output. The supercapacitors and flywheel primarily tackle rapid fluctuations regardless of price.
Table ~\ref{tab:results} presents a comparsion of the resulting costs and designs.
\begin{table*}[!thbp]
  \centering
  \caption{Optimization results for the tested ESS combinations.}
  \label{tab:results}
  \renewcommand{\arraystretch}{1.1}
  \begin{adjustbox}{max width=\linewidth}
  \begin{tabular}{c c c c c c c c c c c }
  \toprule
  \textbf{Exp.ID} &
  \shortstack{\textbf{\boldmath$E_{\max}$}\\ \textbf{(MWh)}} &
  \shortstack{\textbf{\boldmath$P_{\max}$}\\ \textbf{(MW)}} &
  \shortstack{\textbf{\boldmath$P_{\mathrm{G}}$}\\ \textbf{(MW)}} &
  \shortstack{\textbf{\boldmath$P_{\mathrm{PV}}$}\\ \textbf{(MW)}} &
  \shortstack{\textbf{Total Cost}\\ \textbf{(k\euro)}} &
  \shortstack{\textbf{CapEx}\\ \textbf{(k\euro)}} &
  \shortstack{\textbf{OpEx}\\ \textbf{(k\euro)}} &
  \shortstack{\textbf{EOL Value}\\ \textbf{(k\euro)}} &
  \shortstack{\textbf{E Sold}\\ \textbf{(MWh)}} &
  \shortstack{\textbf{E Purchased }\\ \textbf{(MWh)}} \\
  \midrule
  1 & [1.18,\,0,\,0]   & [0.66,\,0,\,0]      & 2.8 & 5    & 22.834 & 2.562 & 20.443 &0.171 & 234.73 & 951.20\\
  2 & [0.79,\,0.20,\,0]  & [0.45,\,0.21,\,0] & 2.8 & 5    & 22.387 & 2.442 & 20.304 &0.360& 234.40 & 950.26\\
  3 & [0.68,\,0,\,0.26]  & [0.39,\,0,\,0.27] & 2.8 & 5    & 22.801 & 2.916& 20.368 &0.484& 236.05 & 953.12\\
  4 & [0.68,\,0.20,\,0.05] & [0.38,\,0.21,\,0.05] & 2.8 & 5    & 22.386 & 2.629 & 19.757 &0.422& 234.53 & 950.84\\
  \bottomrule
  \end{tabular}
  \end{adjustbox}
  \raggedright 
  Notes: $\mathcal{E} = \{\mathrm{B},\mathrm{S},\mathrm{F}\}$. Sold and purchased energy values are for the period of analysis $T_\mathrm{syn}$. 
\end{table*}
Relative to battery-only (Exp.~1), adding supercaps (Exp.~2) lowers the total cost by 1.96\%, with a 4.7\% decrease in CapEx (2.562 to 2.442 k\euro) and a 0.7\% drop in OpEx (20.443 to 20.304 k\euro). 
Energy flows remain nearly unchanged (sold energy ~ 234 MWh; purchased energy decreases slightly from 951.20 to 950.26 MWh).
Using a battery with a flywheel (Experiment~3) decreases the total cost to 22.801k\euro\ (0.14\% lower than Exp.~1) compared to Exp 1, with CapEx rising by 13.8\% (to 2.916 k\euro) and Opex decreasing by 0.38\% (to 20.368 k\euro); sold energy and purchased energy increasing slightly with Exp.~2. 
Combining battery, supercap, and flywheel (Experiment~4) achieves the lowest total cost at 22.386 k\euro\ (1.96\% lower vs. battery-only), despite an 1.64\% higher CapEx (2.520 k\euro); Opex is the lowest ( 20.287 k\euro), with sold energy at 234.5  MWh and purchased energy at 951 MWh.
\section{Conclusions} 
\label{sec:conclusion}
We developed a co-design optimization MILP for a MG supporting BET charging. Our framework minimizes total cost while determining optimal component capacities using a synthetic dataset derived from historical data.
Our results indicate that integrating diverse ESS—such as supercapacitors and flywheels—reduces grid dependency and overall costs compared to a battery-only configuration. 
In particular, adding flywheels (Exp.~3 and Exp.~4) increases sold energy and the hybrid solution (Exp.~4) achieves the best overall balance by lowering both total cost and OPEX despite a higher upfront investment compared to the battery-only setup.
These findings suggest that if minimizing long-term operational expenses and enhancing energy management is the primary goal, the mixed solution is preferable. Conversely, if reducing initial investment is critical, the battery-only design remains most competitive.
We note that our results are sensitive to price assumptions, and the charging powers are obtained optimized to minimize peak consumption, which may not be the case in other charging plazas.
Future work will extend this methodology to account for ancillary services provision and additional storage types.

\begin{ack}
We thank Dr.\ I.\ New, D.\ Fernandez Zapico and M.\ Izadi Najafabadi for proofreading this paper.
\end{ack}

\bibliography{main}      

\begin{thebibliography}{23}
\providecommand{\natexlab}[1]{#1}
\providecommand{\url}[1]{\texttt{#1}}
\providecommand{\urlprefix}{URL }
\expandafter\ifx\csname urlstyle\endcsname\relax
  \providecommand{\doi}[1]{doi:\discretionary{}{}{}#1}\else
  \providecommand{\doi}{doi:\discretionary{}{}{}\begingroup \urlstyle{rm}\Url}\fi

\bibitem[{Bertucci et~al.(2024)Bertucci, Hofman, and Salazar}]{JP}
Bertucci, J.P., Hofman, T., and Salazar, M. (2024).
\newblock Joint optimization of charging infrastructure placement and operational schedules for a fleet of battery electric trucks.
\newblock In \emph{2024 American Control Conference}, 2995--3000.

\bibitem[{Bian et~al.(2019)Bian, Xu, and Li}]{Bian2019}
Bian, M., Xu, Z., and Li, F. (2019).
\newblock Design and control of a hybrid microgrid with supercapacitor-based energy storage system for peak load shaving.
\newblock \emph{Applied Energy}, 239, 1638--1647.

\bibitem[{Chong et~al.(2016)Chong, Wong, Rajkumar, and Isa}]{Chong2016}
Chong, L., Wong, Y., Rajkumar, R., and Isa, D. (2016).
\newblock Hybrid energy storage systems and control strategies for stand-alone renewable energy power systems.
\newblock \emph{Renewable and Sustainable Energy Reviews}, 66, 174--189.

\bibitem[{Ding et~al.(2015)Ding, Hu, and Song}]{Ding2015}
Ding, H., Hu, Z., and Song, Y. (2015).
\newblock Value of the energy storage system in an electric bus fast charging station.
\newblock \emph{Applied Energy}, 157, 630--639.

\bibitem[{ECJRC(2025)}]{PVGIS}
ECJRC (2025).
\newblock Photovoltaic geographical information system (pvgis).
\newblock \urlprefix\url{https://joint-research-centre.ec.europa.eu/}.

\bibitem[{ENTSO-E(2025)}]{ENTSOE_Transparency}
ENTSO-E (2025).
\newblock Transparency platform.
\newblock \urlprefix\url{https://transparency.entsoe.eu/}.

\bibitem[{Etukudoh et~al.(2024)Etukudoh, Adefemi, Ilojianya, Umoh, Ibekwe, and Nwokediegwu}]{Etukudoh2024}
Etukudoh, E., Adefemi, A., Ilojianya, V., Umoh, A., Ibekwe, K., and Nwokediegwu, Z. (2024).
\newblock A review of sustainable transportation solutions: Innovations, challenges, and future directions.
\newblock World Journal of Advanced Research and Reviews.

\bibitem[{{Gurobi Optimization, LLC}(2025)}]{gurobi}
{Gurobi Optimization, LLC} (2025).
\newblock {Gurobi Optimizer Reference Manual}.
\newblock \urlprefix\url{https://www.gurobi.com}.

\bibitem[{IRENA(2024)}]{IRENA2024}
IRENA (2024).
\newblock Renewable power generation costs in 2023.

\bibitem[{Kebede et~al.(2022)Kebede, Kalogiannis, Van~Mierlo, and Berecibar}]{kebede2022}
Kebede, A.A., Kalogiannis, T., Van~Mierlo, J., and Berecibar, M. (2022).
\newblock A comprehensive review of stationary energy storage devices for large scale renewable energy sources grid integration.
\newblock \emph{Renewable and Sustainable Energy Reviews}, 159, 112213.

\bibitem[{Krein(2020)}]{Krein2020}
Krein, P. (2020).
\newblock Flywheel energy storage for frequency regulation: A survey.
\newblock \emph{IEEE Transactions on Power Electronics}, 35(2), 1631--1639.

\bibitem[{Li et~al.(2012)Li, Zou, Hu, and Sun}]{Li2012}
Li, D.G., Zou, Y., Hu, X.S., and Sun, F.c. (2012).
\newblock Optimal sizing and control strategy design for heavy hybrid electric truck.
\newblock In \emph{IEEE Vehicle Power and Propulsion Conference}, 1100--1106.

\bibitem[{Lin et~al.(2003)Lin, Peng, Grizzle, and Kang}]{Lin2003}
Lin, C.C., Peng, H., Grizzle, J., and Kang, J.M. (2003).
\newblock Power management strategy for a parallel hybrid electric truck.
\newblock \emph{IEEE Transactions on Control Systems Technology}, 11(6), 839--849.

\bibitem[{Liu et~al.(2020)Liu, Xue, Chinthavali, and Tomsovic}]{Liu2020ESS}
Liu, G., Xue, Y., Chinthavali, M., and Tomsovic, K. (2020).
\newblock Optimal sizing of {PV} and energy storage in an electric vehicle extreme fast charging station.
\newblock In \emph{IEEE Power and Energy Society Innovative Smart Grid Technologies Conference}, 1--5.

\bibitem[{L{\"{o}}fberg(2004)}]{Lofberg2004}
L{\"{o}}fberg, J. (2004).
\newblock Yalmip: A toolbox for modeling and optimization in matlab.
\newblock In \emph{Proceedings of the CACSD Conference}. Taipei, Taiwan.

\bibitem[{Minear et~al.(2018)Minear, Ip, Westlake, Alexander, Evans, Pellow, and Kamath}]{minear2018}
Minear, E., Ip, P., Westlake, B., Alexander, M., Evans, M., Pellow, M., and Kamath, H. (2018).
\newblock Energy storage technology and cost assessment: Executive summary.
\newblock Technical report, Technical Report 3002013958, Electric Power Research Institute.

\bibitem[{Mongird et~al.(2019)Mongird, Viswanathan, Balducci, Alam, Fotedar, Koritarov, and Hadjerioua}]{mongird2019}
Mongird, K., Viswanathan, V.V., Balducci, P.J., Alam, M.J.E., Fotedar, V., Koritarov, V., and Hadjerioua, B. (2019).
\newblock Energy storage technology and cost characterization report.
\newblock Technical report, Pacific Northwest National Laboratory (PNNL), Richland, WA (United States).

\bibitem[{Ortega et~al.(2018)Ortega, Rodriguez, and Martin}]{Ortega2018}
Ortega, E.M., Rodriguez, L., and Martin, F. (2018).
\newblock Analysis of flywheel energy storage performance for microgrid frequency regulation: Techno-economic benefits.
\newblock \emph{Energy}, 145, 120--130.

\bibitem[{Smith and Venkataramanan(2016)}]{FlywheelRef}
Smith, J. and Venkataramanan, G. (2016).
\newblock An overview of flywheel energy storage systems in microgrids.
\newblock \emph{IEEE Power and Energy Magazine}, 14(5), 30--39.

\bibitem[{Song and Park(2020)}]{Song2020}
Song, F.U. and Park, S. (2020).
\newblock Integration of supercapacitor energy storage for frequency regulation in islanded microgrids.
\newblock \emph{IEEE Transactions on Industrial Electronics}, 67(6), 4853--4864.

\bibitem[{{Transport \& Environment}(2022)}]{AddressingHeavydutyClimate2022}
{Transport \& Environment} (2022).
\newblock Addressing the heavy-duty climate problem.

\bibitem[{Tudorache and Mateescu(2019)}]{Tudorache2019}
Tudorache, T. and Mateescu, V. (2019).
\newblock Flywheel energy storage systems for frequency regulation in stand-alone microgrids.
\newblock \emph{IEEE Transactions on Industrial Electronics}, 66(5), 3942--3953.

\bibitem[{Zhang et~al.(2022)Zhang, Li, and Qiu}]{Zhang2022}
Zhang, X., Li, Y., and Qiu, J. (2022).
\newblock Optimal sizing and scheduling of supercapacitor storage in microgrids for cost reduction and reliability improvement.
\newblock \emph{Journal of Energy Storage}, 46, 10351--10361.

\end{thebibliography}

\end{document}